\documentstyle[twocolumn,prl,aps,graphicx]{revtex}
\tightenlines
\begin{document}
\draft
\title{Hartree-Fock dynamics in highly excited quantum dots}
\author{Antonio Puente, Lloren\c{c} Serra}
\address{Departament de F\'{\i}sica, Universitat de les Illes Balears,
E-07071 Palma de Mallorca, Spain}
\author{Vidar Gudmundsson}
\address{Science Institute, University of Iceland, Dunhaga 3, 
IS-107 Reykjavik, Iceland}
\date{June 7, 2001}
\maketitle
\begin{abstract}
Time-dependent Hartree-Fock theory is used to describe density 
oscillations of symmetry-unrestricted two-dimensional nanostructures. 
In the small amplitude limit the results reproduce those obtained within 
a perturbative approach such as the linearized time-dependent  
Hartree-Fock one. The nonlinear regime is explored by studying
large amplitude oscillations in a non-parabolic 
potential, which are shown to introduce a strong coupling with internal 
degrees of freedom. This excitation of internal modes, mainly of monopole 
and quadrupole character, results in sizeable modifications of the dipole 
absorption.
\end{abstract}

Hartree-Fock (HF) theory is one of the most successful microscopic 
schemes to address finite Fermi systems, and usually the starting 
point of more refined approximations. Shortly after its proposal   
Dirac suggested a time-dependent extension (tdHF) that would account 
for dynamical processes \cite{Dir30} which has been the basis of 
many studies of oscillations in systems such as atoms, 
nuclei and molecules. Time-dependent HF theory is not restricted to 
small amplitude oscillations, although in this limit one obtains 
a scheme based on ground state perturbations which is not 
explicitly time dependent. In fact, one can obtain
linearized equations (tdHF*) which are essentially equivalent to the 
well-known random-phase approximation (RPA) \cite{Blaxx}.

In this Letter we numerically study the tdHF
evolution for electrons confined in a two-dimensional semiconductor 
quantum dot by direct integration of the time-dependent equations. 
We show the equivalence with the linearized tdHF* approach previously used by 
one of us \cite{Gudxx} and, at the same time, prove the minor role of the ground 
state symmetry-breaking on the far-infrared (FIR) absorption. 
We also explore the nonlinear regime by going to the large
amplitude limit where we find 
a fast energy transfer from dipole to internal breathing and 
quadrupole oscillations, made possible by the non-parabolicity 
of the confining  potential. The excitation of internal breathing 
and quadrupole modes results in sizeable modifications
of the dipole absorption, as compared to the small amplitude case.

As described in many textbooks, the essential ingredient of tdHF theory 
is the description of the state of the system at 
any time of the dynamic evolution by a single Slater determinant, 
built with single-particle orbitals. Given initial conditions, 
the central equations of the theory provide the time evolution  
of the orbital set $\{\varphi_{i\eta}({\bf r},t)\}$. 
The orbital labels indicate state number ($i$) and spin 
$z$-projection ($\eta=\uparrow, \downarrow$). 
The tdHF equations for electrons confined to the $xy$ plane by an external 
$v_{ext}({\bf r},t)$ and vector 
${\bf A}({\bf r})=B/2(-y,x)$ potentials read \cite{units}
\begin{eqnarray}
      \label{eq2p1}
      i\frac{\partial}{\partial t}\varphi_{i\eta}({\bf r}_1,t) &=&
      \left[\rule{0cm}{0.6cm}
      \frac{1}{2} \left(-i\nabla+\gamma {\bf A}({\bf r}_1)\right)^2 +
      v_H({\bf r}_1,t) \right.\nonumber\\
      &+& \left.
      v_{ext}({\bf r}_1,t) \, + \frac{1}{2} g^* m^* \gamma B s_z 
      \rule{0cm}{0.6cm}\right] 
      \varphi_{i\eta}({\bf r}_1,t)\nonumber\\
      &-& \int{d{\bf r}_2\, 
      \frac{\rho_\eta({\bf r}_1,{\bf r}_2,t)}{r_{12}}\,
      \varphi_{i\eta}({\bf r}_2,t)}\; , 
\end{eqnarray}
where we have defined $\gamma=e/c$ and the Hartree potential
$v_H({\bf r}_1,t)=\int{d{\bf r}_2\, \rho({\bf r}_2,t)/r_{12}}$,
given in terms of the total density $\rho({\bf r},t)$. In Eq.\ (\ref{eq2p1})
we have also defined the one-body density-matrix entering the exchange integral
\begin{equation}
      \rho_\eta({\bf r}_1,{\bf r}_2,t) =
      \sum_{j,occ.}{\varphi_{j\eta}({\bf r}_1,t) \, 
              \varphi^*_{j\eta}({\bf r}_2,t)
      } \; .
\end{equation}

In this work we shall present direct numerical  
solutions of Eq.\ (\ref{eq2p1})
obtained by discretizing the equations both in space, into a uniform grid of 
points, and in time $t_k = k \Delta t$ by using the Cranck-Nicholson 
algorithm \cite{NR}. Specifically, by
calling $h_0^{(k)}$ and ${\cal V}_{i\eta}^{(k)}$ 
the local term (within square brackets) and the exchange integral  
in Eq.\ (\ref{eq2p1}) at time-step $k$, the algorithm reads
\begin{eqnarray}
      \label{eq2p3}
      \left(
      1+{i\Delta t\over 2} h_0^{(k+1)}
      \right)
      \varphi_{i\eta}^{(k+1)}
      &=&
      \left(
      1-{i\Delta t\over 2} h_0^{(k)}
      \right)
      \varphi_{i\eta}^{(k)}\nonumber\\
      &+&
      {i\Delta t\over 2}
      \left( 
      {\cal V}_{i\eta}^{(k)}+{\cal V}_{i\eta}^{(k+1)}
      \right)\; .
\end{eqnarray} 
As a first guess, we assume $h_0^{(k+1)}=h_0^{(k)}$
and ${\cal V}_{i\eta}^{(k+1)}={\cal V}_{i\eta}^{(k)}$ which allows an 
initial solution of (\ref{eq2p3}). Subsequently  $h_0^{(k+1)}$
and  ${\cal V}_{i\eta}^{(k+1)}$ are updated and the process repeated 
untill convergence in $\varphi_{i\eta}^{(k+1)}$ is found 
(typically 4-8 iterations are enough for a sufficiently small $\Delta t$).

The system strength function $S(\omega)$, or equivalently the absorption 
cross section $\sigma(\omega)=\omega S(\omega)$ are easily determined from the 
time signals $\langle \sum_i{{\bf r}_i}\rangle(t)$ by means of a frequency 
analysis, in a much similar way as used in real-time Density-Functional 
calculations of atomic clusters \cite{Yabxx,Reixx} and quantum 
dots \cite{Pue99}. Important differences with the latter are  
the exact treatment of the exchange (neglecting correlation though)
and the possibility to explore large amplitude oscillations 
since the formalism is not centered on a ground-state energy functional.

In the limit of small oscillation amplitudes we compare with a frequently used 
linearized version (tdHF*), which provides the linear response 
to an external oscillating dipole potential adiabatically switched on
\begin{equation}
      {\phi}^{\mathrm ext}({\bf r},t)={\cal E}^{\mathrm ext}r
      \exp\left[ \pm i\theta -i(\omega +i\eta) t\right] \; ,
\end{equation}
with $(r,\theta)$ the polar
coordinates, $\eta\rightarrow 0^+$ and $`\pm '$ labelling the two circular 
polarization states. The power absorbed by the system is 
\begin{eqnarray}
      P(\omega )&=&2\pi \, \omega{\cal E}^{\mathrm ext}\sum_{\alpha\beta}
      \langle\beta |r|\alpha\rangle \delta_{M_{\beta},{M_{\alpha}\pm 1}}
      \times\nonumber\\
      &&\Im\left\{ f^{\alpha\beta}(\omega)
      \langle\alpha |\phi^{\mathrm sc} |\beta\rangle\right\} , 
\end{eqnarray}
where $\phi^{\mathrm sc}$ is the self-consistent electrostatic potential 
found from the external perturbation $\phi^{\mathrm ext}$ via the 
dielectric tensor \cite{Gudxx}, and
\begin{equation}
      f^{\alpha\beta}(\omega)= 
      \frac{f^0_\beta-f^0_\alpha}{\omega +(\omega_\beta-\omega_\alpha)
      +i\eta} \; ,
\end{equation}
with $f^0$ the equilibrium Fermi distribution. Notice that within the tdHF*
we require the angular part of the orbitals to be simple phases
$e^{iM\theta}$, which amounts to assuming circular symmetry of the 
density and mean field.

Most of the system physical properties are determined by the external
potential $v_{ext}(r)$, which we shall assume to be static and circularly 
symmetric. Following Ref.\ \cite{Kra01} we consider a 
step in the radial direction separating two regions with a 
different slope in the potential. This type of behaviour
reproduces to some 
extent the effective potential in quantum dot arrays, where 
lattice periodicity introduces a deviation from a pure quadratic 
radial law. 
The specific parametrization we use is
\begin{eqnarray}
      \label{eq3p7}
      v_{ext}(r)&=&v_1(r)\, f_1(r)+\left[
      v_2(r)\right.\nonumber\\
      &+&\left. v_1(R_0)-v_2(R_0)+\Delta
      \right]\, f_2(r)\; ,
\end{eqnarray}
where $R_0$ and $\Delta$ are the step position and height, respectively. 
In Eq.\ (\ref{eq3p7}) $v_{1,2}(r)$ indicate the potential 
in the $r<R_0$ ($r>R_0$) region and $f_{1,2}(r)$ is a smooth 
switching function. 
Namely, these region-dependent functions are  
\begin{eqnarray}
      v_{1,2}(r)&=&\frac{1}{2}\omega_{1,2}\, r^2\nonumber\\
      f_{1,2}(r)&=&\frac{1}{2}\left(
      1\mp \tanh{r-R_0\over \sigma }
      \right)\; .
\end{eqnarray} 
In the calculations shown below we consider a system with 
$N=6$ electrons 
and $R_0=30$~nm, $\sigma=5$~nm, $\omega_{1,2}=3.37$~meV, $\Delta=6$~meV. This 
particular set of values is motivated by the experiments in 
Ref.\ \cite{Kra01}. 

Figure 1 displays the dipole absorption in the small amplitude
regime obtained within tdHF and tdHF*.
As is well known from Kohn's theorem, in purely parabolic dots the 
FIR absorption consists of two bands at energies 
\begin{equation}
      \omega_\pm=\sqrt{\omega_0^2+(\omega_c/2)^2}\pm\omega_c/2\;
\end{equation}
where $\omega_c=eB/c$ is the cyclotron frequency. 
Looking at the lower panel of Fig\ 1, one can indeed identify a low energy 
peak 
around 2 meV and a higher-energy structure with two peaks. 
The existence of a structured high 
energy band is in agreement with the experiments \cite{Kra01}
and as seen in Fig.\ 1 the two methods yield very similar
results, showing the 
equivalence of the very different techniques in this limit. 
Quite remarkably, at $B=1$T HF provides a symmetry broken ground 
state, with the six electrons localized in a ring-like structure similar 
to those discussed by Yannouleas and Landman \cite{Yanxx} or 
Manninen {\em et al.} \cite{Manxx}, 
which is obviously not present within the circular HF*. 
Nevertheless, Fig.\ 1 shows that
the absorption obtained from the symmetry-broken 
ground state does not importantly differ from that of the circular solution, 
proving that the details of the FIR absorption are not much sensitive to the 
internal structure. For a purely parabolic dot there is a complete 
independence of the FIR absorption and the internal symmetry but
it is a remarkable fact that for the more realistic dot potential 
(\ref{eq3p7}) the details of the high energy branch are not sensitive 
to the internal structure.

We turn next to large amplitude oscillations within tdHF. Figure 2
shows the center of mass trajectory following an initial rigid displacement 
to the point indicated in the upper right corner. This is a dramatic 
perturbation which amounts to an excitation energy 
of $\Delta E=26.5$ meV, to be compared with a ground state energy 
of $E=112.7$ meV. 
A relevant observation in Fig.\ 2 is that the amplitude of the  
oscillation dampens in time, hinting at an important energy transfer from 
the center of mass to the internal degrees of freedom which are evolving in 
time. To better monitor this coupling mechanism we introduce the 
internal coordinates $\tilde{\bf r}={\bf r}-{\bf R}_{cm}$, where 
${\bf R}_{cm}$ is the center of mass position, and define the {\em internal}
quadrupole and monopole operators as
\begin{eqnarray}
      \tilde{Q} &=& \sum_i{\tilde{x_i}\tilde{y_i}}=
      \sum_i{x_iy_i}-\frac{1}{N}\sum_{ik}{x_iy_k}\nonumber\\
      \tilde{M} &=& \sum_i{\tilde{x_i}^2+\tilde{y_i}^2}=
      \sum_i{x_i^2+y_i^2}-\frac{1}{N}\sum_{ik}{x_ix_k+y_iy_k}\; .
\end{eqnarray}
Notice that these internal probes involve two-body operators when 
expressed in the fixed laboratory frame.

Figure 3 shows the time signals corresponding to the large amplitude 
displacement of Fig.\ 2 as well as the results following a much 
smaller perturbation corresponding to the linear regime. As expected 
from the previous figure, 
the large oscillation dipole signal is reducing its amplitude   
in time, while the internal quadrupole and monopole modes get
highly excited. The breathing-mode result is actually indicating that 
the dot expands after a 
rather short time, and keeps oscillating around the expanded configuration.
This energy transfer to the breathing mode is not seen in the small amplitude
regime, where only minor oscillations around the ground state configuration 
are always found. As already mentioned, the mechanism observed here is possible 
due to the non-parabolic confining potential permitting the energy 
transfer between modes, showing that
nonlinear processes associated with high energy excitations can effectively 
induce strong internal excitations.

The frequency analysis of the large amplitude signal is complicated 
by the decrease in amplitude. An approximate treatment is still possible 
by defining a time window in which the variations are not 
too large. In fact the frequency spectrum evolves smoothly in time, 
depending on the window position. For large times the 
result is not much sensitive on this position and is shown in Fig.\ 4 
for $B=1$T. 
Notice that in the frequency analysis of the time signal each peak 
is represented by a Lorentzian having an artificially chosen width 
$\Gamma$, for an easier representation of the spectrum.
Although experimentally rather large 
$\Gamma$'s are usually found, in Fig.\ 4 we show the result obtained 
using either a large or a small 
value of $\Gamma$, in order to clearly display the fine structure of 
the absorption. 

An important fragmentation appears 
in the large amplitude spectrum of Fig.\ 4, due the splitting in many 
closely lying peaks. Actually, the three upper sharp transitions
of the small amplitude spectrum transform into a bunch of peaks.
Within a one-particle picture, we attribute this 
effect to the large fluctuation of the mean field, seen from the internal 
signals, which causes also important variations in the effective single 
particle energies and in the corresponding particle-hole transitions.
Interestingly enough, when using a large (and more realistic) $\Gamma$
we obtain a narrowing of the high energy branch, roughly covering
the interval $[4.75,5.5]$~eV. In the small amplitude case the higher
branch, composed of three peaks, approximately lies at $[4.25,5.5]$~eV.
The observed effect is thus a {\em dynamical narrowing} of the high
energy branch induced by the large amplitude oscillation.

To summarize, the tdHF equations have been integrated in time to 
describe electronic oscillations in two-dimensional quantum dots
confined by a circular non-parabolic potential. 
In the limit of small amplitudes the results are very similar to 
those obtained in the linearized version of the circularly constrained
theory (tdHF*). The symmetry broken ground states do not significantly 
modify the dipole absorption in this regime. 
The nonlinear behaviour has been investigated by using large amplitude 
initial displacements. An effective transfer from the center of mass dipole 
to internal monopole and quadrupole modes has been observed in this case, 
thus providing a mechanism to strongly excite internal modes
in non-parabolic dots. Additionally, the large amplitude motion 
has been found to
induce a dynamical narrowing of the dipole absorption. This 
effect would manifest in absorption experiments using a high-intensity
far-infrared radiation. 
 
This work was supported in part by Grant No.\ PB98-0124 from DGESeiC, 
Spain, the Research Fund of the University of Iceland, and the Icelandic
Natural Science Council.

\begin{figure}
\begin{center}
      \includegraphics[clip,width=8cm]{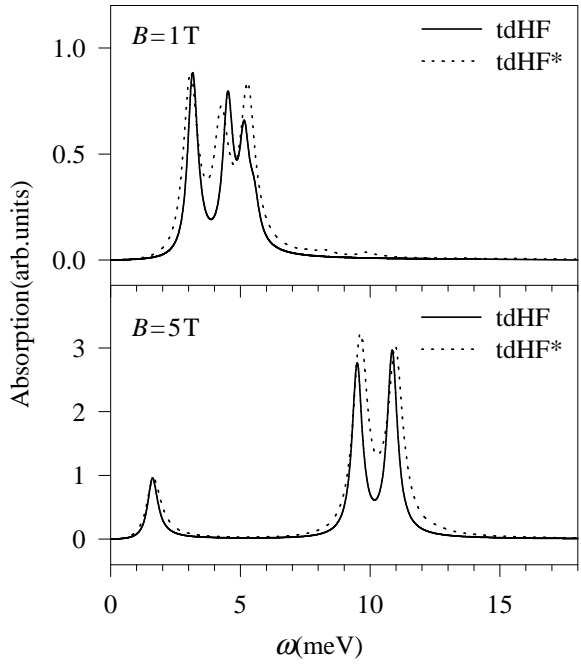}
\end{center}
\caption{The power absorption for 6 electrons obtained 
within tdHF and using the linear version 
with circular symmetric tdHF*, for two different values of the magnetic field 
$B$.}
\label{ComPL}
\end{figure}

\begin{figure}
\begin{center}
      \includegraphics[clip,width=7cm]{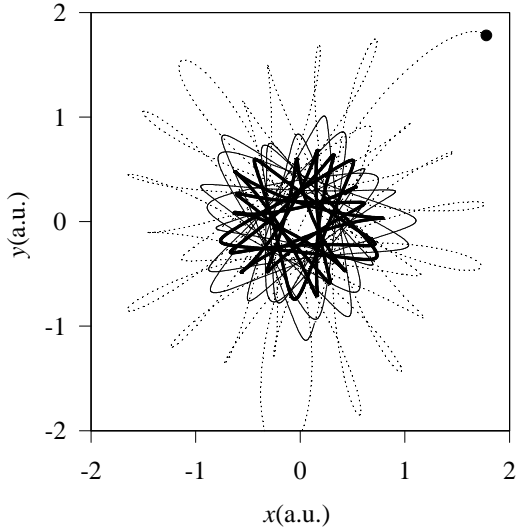}
\end{center}
\caption{Center of mass trajectory following a large initial displacement to the 
point in the upper rigth corner, obtained within tdHF for the quantum dot 
considered in this work at $B=1$T. Dotted, thin solid and thick solid lines 
indicate the motion in subsequent intervals of 
$\approx 12$ ps, respectively. See text.}
\label{CM}
\end{figure}

\begin{figure}
\begin{center}
\includegraphics[clip,width=8.5cm]{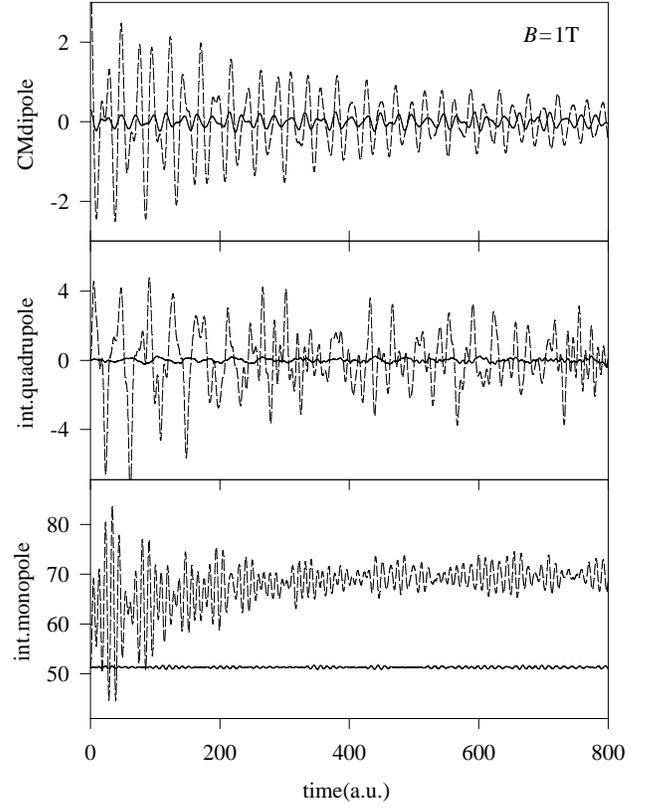} 
\end{center}
\caption{Time dependence of the center of mass (CM) dipole, internal quadrupole
and internal monopole signals in the small (solid) and large (dashed) amplitude 
regimes at $B=1$~T.}
\label{tsigns}
\end{figure}

\begin{figure}
\begin{center}
      \includegraphics[clip,width=8.5cm]{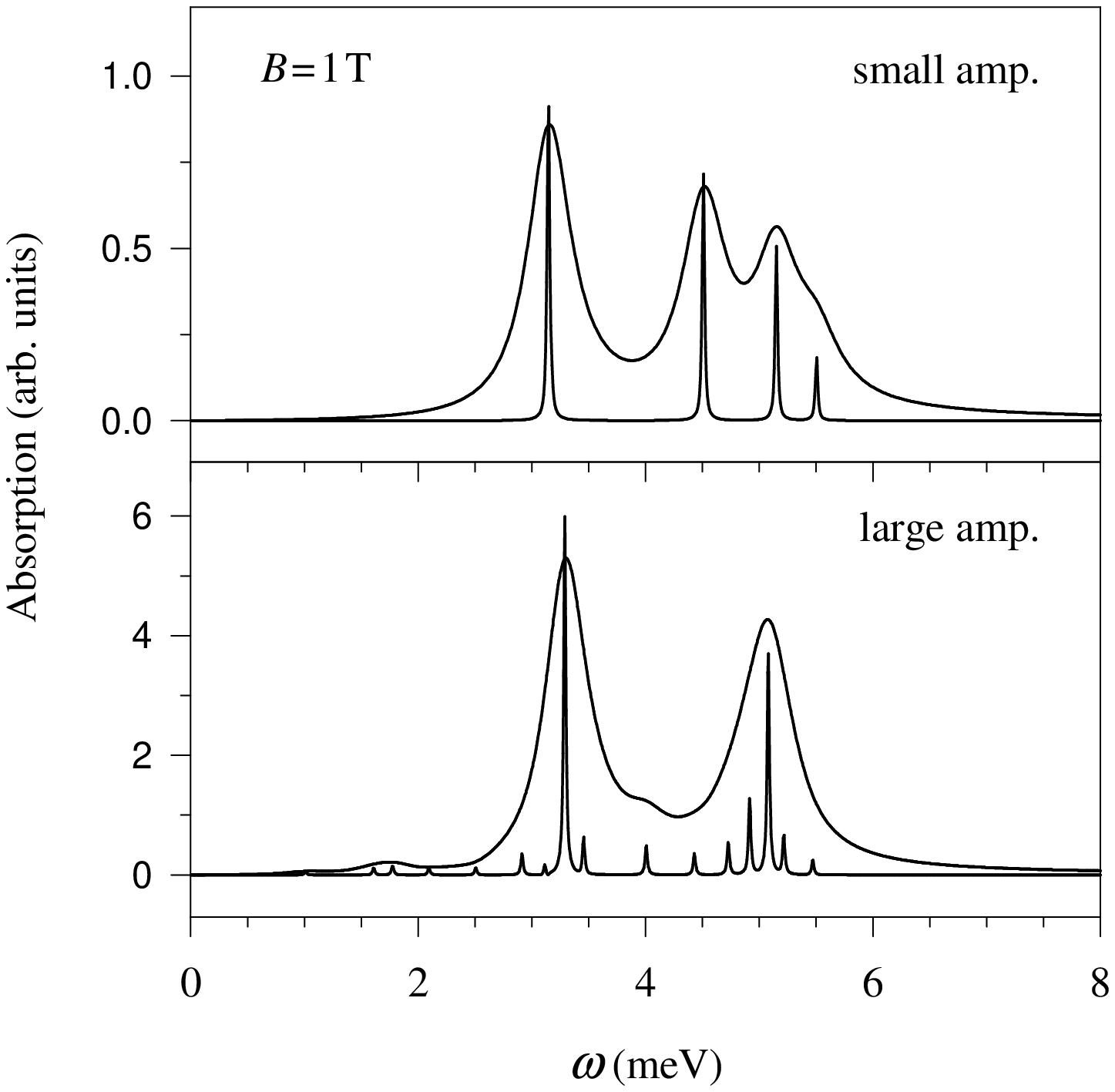}
\end{center}
\caption{Dipole absorption corresponding to the small (upper panel) and 
large (lower panel) amplitude signals of Fig.\ 3. 
Two different peak widths have been used for a better presentation. 
The large amplitude results are obtained from the time window [500,800] a.u.}
\label{spec4}
\end{figure}

\end{document}